\documentclass[10pt]{article}

\newif\ifeng
\engtrue

\ifeng
\else
\usepackage{hangul}
\fi
\usepackage{epsfig}
\usepackage{amssymb}
\usepackage{amsmath}
\usepackage{amsthm}
\usepackage{comment}
\usepackage{boxedminipage}
\usepackage{verbatim}
\usepackage{url}
\usepackage{array}
\usepackage{color}
\usepackage{soul}

\begin{document}

%\vspace*{-1.5cm}\hspace*{10.0cm} Date: 21 Feb. 2014 (rev. 6)

\ifeng
\else

%%%%%%%%%%%%%%%%  ---- Title ---- %%%%%%%%%%%%%%%%%a
\pagestyle{empty}

\vspace*{2.0cm}

\begin{center}
{\LARGE \bf Odysseus/DFS: Integration of DBMS and Distributed File System
for Transaction Processing of Big Data} \\
\vspace*{0.5cm}

{\LARGE \bf 오디세우스/DFS: Big Data의 트랜잭션 처리를 위한 DBMS와 DFS의 결합}
\vspace*{1.0cm}

{\large 김준성, 황규영, 권혁윤} \\
{\large Jun-Sung Kim, Kyu-Young Whang, Hyuk-Yoon Kwon}
\vspace*{1.0cm}

{\large 한국과학기술원 전산학과} \\
{\normalsize Department of Computer Science} \\
{\normalsize Korea Advanced Institute of Science and Technology\,(KAIST)} \\

\end{center}

\vspace*{2.1cm}

\hspace*{1.50cm}{연구(투고) 분야: 데이터베이스} \\
\hspace*{2.28cm}

\hspace*{1.50cm}{전화: (042)350-7722/7822} \\
\hspace*{2.28cm}{Fax: (042)867-3562} \\
\hspace*{2.28cm}{e-mail: \{jskim, kywhang, hykwon\}@mozart.kaist.ac.kr} \\

%%%%%%%%%%%%%%%%%%%%%%%%%%%%%%%%%%%%%
\newpage
\vspace*{-1.50cm}
\fi

\pagestyle{plain}
\pagenumbering{arabic}
\setcounter{page}{1}

\vspace*{-0.10cm}
\begin{center}
{\LARGE \bf Odysseus/DFS: Integration of DBMS and Distributed File System
for Transaction Processing of Big Data} \\
\vspace*{0.20cm}

\ifeng
Jun-Sung Kim$^{1}$, Kyu-Young Whang$^{1}$, Hyuk-Yoon Kwon$^{1}$, and Il-Yeol Song$^{2}$ \\
\vspace*{-0.25cm}
$^{1}$ Department of Computer Science, KAIST, Daejeon, Korea, \\
\vspace*{-0.35cm}
\{jskim, kywhang, hykwon\}@mozart.kaist.ac.kr \\
\vspace*{-0.25cm}
$^{2}$ College of Information Science and Technology, Drexel University, Philadelphia, USA, \\
\vspace*{-0.35cm}
songiy@drexel.edu \\
\else {\LARGE \bf 오디세우스/DFS: Big Data의 트랜잭션 처리를 위한
DBMS와 DFS의 결합} \fi

\end{center}
\vspace*{-0.35cm}

\begin{abstract}
{\small
The relational DBMS (RDBMS) has been widely used
since it supports various high-level functionalities such as SQL, schemas, indexes,
and transactions that do not exist in the O/S file system. But, a
recent advent of {\em big data} technology facilitates development
of new systems that sacrifice the DBMS functionality in order to
efficiently manage large-scale data. Those so-called NoSQL systems use
a distributed file system, which support scalability and
reliability. They support scalability of the system by storing
data into a large number of low-cost commodity hardware and
support reliability by storing the data in replica. However, they
have a drawback that they do not adequately support high-level DBMS functionality.
In this paper, we propose
an architecture of a DBMS that uses the DFS as storage. With
this novel architecture, the DBMS is capable of supporting
scalability and reliability of the DFS as well
as high-level functionality of DBMS.
Thus, a DBMS can utilize a virtually unlimited storage space
provided by the DFS, rendering it to be suitable for big data analytics.
As part of the architecture of the DBMS, we propose the notion of the meta
DFS file, which allows the DBMS to use the DFS as the
storage, and an efficient transaction management method including
recovery and concurrency control.
We implement this architecture in Odysseus/DFS, an integration
of the Odysseus relational DBMS\,{\cite{Wha14}}, that has been being
developed at KAIST for over 24 years, with the DFS.
Our experiments on transaction
processing show that, due to the high-level functionality of
Odysseus/DFS, it outperforms Hbase, which is a representative
open-source NoSQL system. We also show that, compared with
an RDBMS with local storage, the performance of Odysseus/DFS is
comparable or marginally degraded, showing that the overhead of
Odysseus/DFS for supporting scalability by using the DFS as the
storage is not significant.
}
\end{abstract}

\newtheorem{definition}{Definition}
\newtheorem{lemma}{\bf Lemma}
\newtheorem{theorem}{\bf Theorem}
\newtheorem{algorithm}{\bf Algorithm}
\newtheorem{property}{\bf Property}
\newtheorem{corollary}{\bf Corollary}

\theoremstyle{definition}
\newtheorem{example}{Example}

\section{Introduction}
\label{ch:Introduction}

\subsection{Necessities of Large-Scale Data Management}

\label{ch:Int1} The relational database management system (RDBMS)
has been widely used as an optimal solution to systemically manage
structured data for decades\,\cite{Sto07}. The RDBMS provides
improved productivity for both producers and users of the
application programs. Specifically, compared to the low-level
functionalities of the O/S file system, the RDBMS supports various
high-level functionalities such as data types, schemas,
transactions, indexes, and SQL.

Rapid advancement of computing systems and computer networks
facilitated numerous services and applications that did not exist before,
and the necessity for managing large-scale data has naturally emerged
during this process\,\cite{GR11}. Explosive growth of large-scale
digital data from social media, Web, sensors, and other data sources
made it impossible to manage them efficiently by simply extending
the current computing paradigm. The term {\em big data} was coined
to cope with such problems of explosive inflation of data, which
need to be solved as a new area of computing\,\cite{Dig12}. In
practice, it was reported that the digital information content of
the world amounted to 1.8 zettabytes\footnote{Zettabytes are a million
petabytes.} in 2011 and was to increase by tens to hundreds of
times in ten years\,\cite{GR11}. It was reported that Facebook
currently manages over 140 billion photos, and more than 7 billion
photos are uploaded each month\,\cite{Ale12}. It means that a single
Internet service manages data over 100 petabytes, assuming the size
of a photo is about one megabyte.

\subsection{Techniques for Large-Scale Data Management}

As the amount of data that computer systems need to manage increases, a
number of techniques for managing
large-scale data were developed\,\cite{Aba09,ABA+09,OG10,Pav09,Val93,Waa09}.
They can be classified into three categories according to the way the
system is configured.

First, the most basic way is to configure a single node RDBMS with a
massive storage device. In this method, we implement an RDBMS on a
high-performance single machine that can handle a large number of
queries and manage large-scale data using massive storage devices
such as a disk array or a SAN. Generally, this method is widely used
in small-scale companies with small-sized data to manage and has an
advantage that the system configuration is relatively easy. However,
there are physical limitations in configuring beyond a certain level
of storage capacity or providing services beyond a certain level of
transaction loads in a single machine. Capital burden is also a
problem since the price of the massive storage hardware is very
expensive and guaranteeing high level fault-tolerance at the
hardware level is also very expensive.

Second, we can use a distributed/parallel DBMS that
distributes data and processes queries over multiple
nodes\,\cite{Sto10}. This configuration extends capacities of data
storage and query processing by storing data over multiple nodes
and processing queries in parallel. But, the distributed/parallel
DBMS has several drawbacks\,\cite{ABA+09}. (1) System
configuration is complex, (2) software of a distributed/parallel
DBMS is expensive, (3) as the number of nodes increases more than
several tens of nodes, the internode communication overhead
increases exponentially, (4) the architecture has an operability
problem caused by decrease of mean-time-before-failure of the system
due to failure of individual nodes. So, this architecture is
difficult to scale beyond a certain number of nodes.

Third, we can use NoSQL systems. By storing large-scale data in
a distributed file system and processing them in parallel using more
than thousands of commodity machines, NoSQL systems make up for
the drawbacks of the two former architectures\,\cite{Pav09}. NoSQL
systems such as GFS or Hadoop DFS (HDFS) utilize a number of
machines connected through a network and
support scalability, reliability, and availability.
Building
a large-scale storage using NoSQL systems is widely accepted
for big data management since its cost is several times less than
that of using a single-node RDBMS with a disk array or using
a distributed/parallel DBMS\,\cite{Har11}.

\subsection{Motivation}

The distributed file system (DFS) and key-value store are
representative examples of NoSQL systems.
The DFS is designed to store large-scale
data on a distributed network that consists of a huge number of
commodity hardwares. It stores data by the unit of a large block
distributed over those machines\,\cite{GGL03}. The DFS supports scalability
by adding any number of new nodes and fault-tolerance by maintaining
multiple replicas of data blocks.
The key-value store is a storage system for storing
a table-like structure by the unit of $<$key, value$>$ pair\,\cite{CDG+06}.
The key-value store supports scalability and fault-tolerance
by using the DFS as the storage.

\begin{figure}[htb]
\vspace*{0.3cm}
 \centering
 \includegraphics[width=0.78\columnwidth]{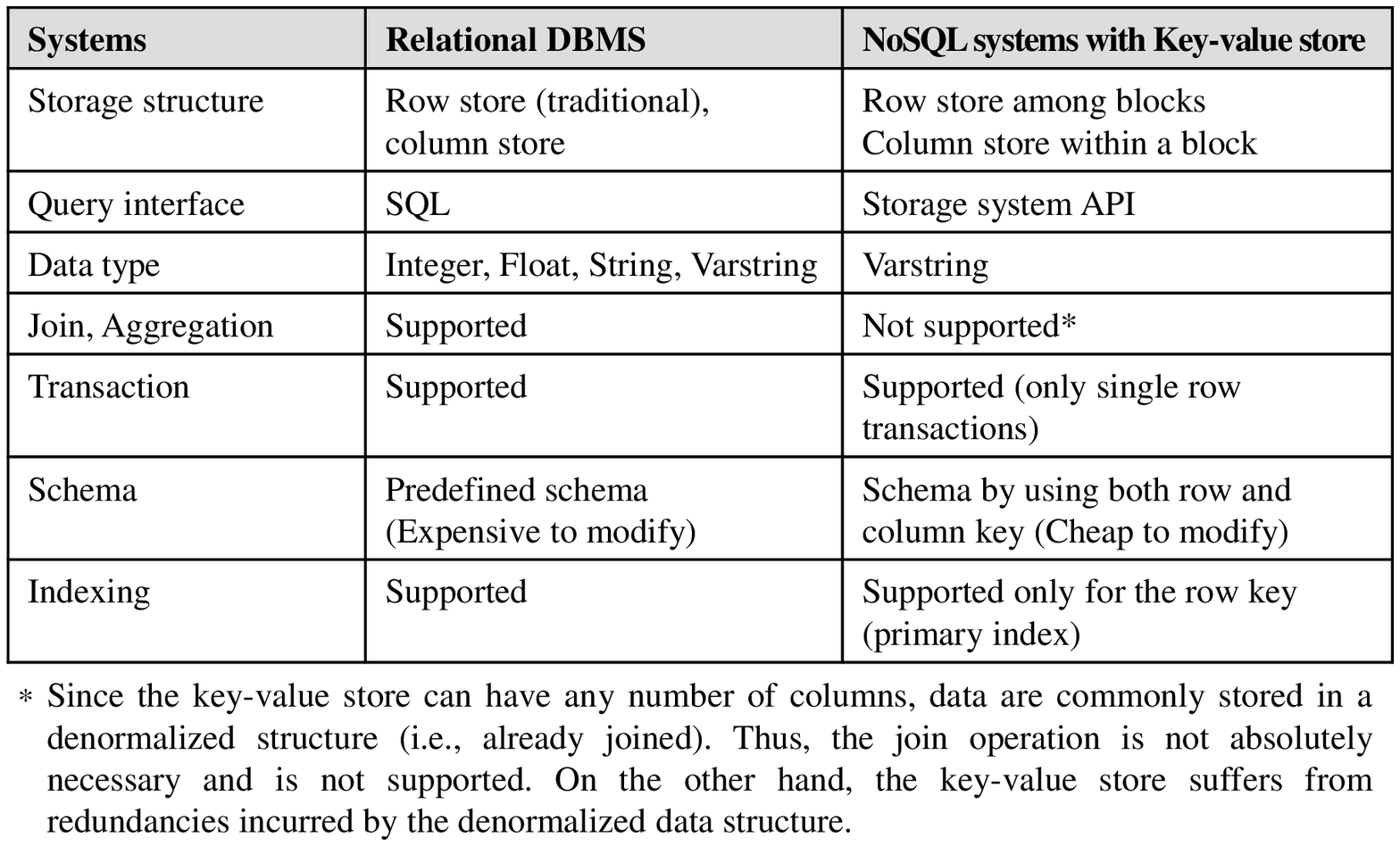}
 \vspace*{-0.6cm}
 \caption{Comparisons of functionalities between the RDBMS and the key-value store.}
 \label{fig:comp1}
 \vspace*{-0.1cm}
\end{figure}

On the other hand, NoSQL systems have disadvantages of lacking
high-level DBMS functionalities compared to the
RDBMS\,\cite{Aba09}. That is, NoSQL systems are designed to
support scalability for large-scale systems by limiting certain
DBMS functionality\,\cite{DG04, GGL03}. Figure \ref{fig:comp1}
compares functionalities of the RDBMS to those of the NoSQL
systems using the key-value store.

There have been a lot of effort to support high-level DBMS
functionality in a NoSQL system\,\cite{ABA+09, ORS+08,
Shu12, TSJ+09}. However, most of those efforts just add some DBMS
functionality to NoSQL systems rather than supporting full
DBMS functionality. We discuss these works in more detail in
Section \ref{ch:RW2}.

\subsection{Our Contributions}

In this paper, we propose a new DBMS architecture that can efficiently manage large
scale big data with full DBMS functionality. We make the following
four major contributions: First, we propose a new architecture, {\em Odysseus/DFS}, that
integrates the RDBMS with the DFS.
Effectively, it is an RDBMS that
uses the DFS as the storage system.
Thus, it supports scalability and fault-tolerance of the DFS
and provides high-level functionalities of the RDBMS.

Second, we propose the notion of the \emph{meta DFS file}, which
allows the RDBMS to use the DFS as the storage. The DFS is a
write-once-read-many storage that do not allow in-place overwriting
and appending. Our
storage manager solves those problems using the notion of the meta DFS file.
The storage manager also performs mapping between a DFS block and DBMS pages.

Third, we propose an efficient transaction management method
including concurrency control and recovery when using the DFS as the storage.
Our method supports schedules serial with respect to write
transactions but concurrent with respect to read transactions. It
supports coarse-granularity locking in order to minimize the
locking overhead in a distributed environment. We note that the
general architecture of Odysseus/DFS is RDBMS-independent. Our
storage management and transaction management methods can be
easily adapted to any RDBMS that uses the page as the unit of I/O.

Last, we show by experiments that, in terms of performance,
Odysseus/DFS outperforms Hbase and  is comparable to or marginally
degraded from the RDBMS. These results are very promising
in that  our methods have a more complete DBMS functionality compared to
Hbase, and yet, have better scalability and fault-tolerance compared to RDBMS.

The rest of this paper is organized as follows. In Section \ref{ch:RelatedWorks},
we review the representative platforms for big data management.
In Section \ref{ch:OdysseusDFS}, we present our new system, Odysseus/DFS.
In Section \ref{ch:Recovery} we present the recovery methods for Odysseus/DFS.
In Section \ref{ch:CC} we discuss how we implement locks for Odysseus/DFS. \,
In Section \ref{ch:Performance}, we
present the performance of Odysseus/DFS compared to other data management systems.
Finally, in Section \ref{ch:Conclusions}, we conclude the paper.

\section{Related Work}
\label{ch:RelatedWorks}

In this section, we review related work that are used for
large-scale data. Section \ref{ch:RW0} reviews recent research
efforts on improving scalability of parallel DBMSs. Section
\ref{ch:RW1} covers NoSQL systems and recent advances on
them. Section \ref{ch:RW2} reviews research efforts on providing
NoSQL systems with high-level DBMS functionality.

\subsection{Parallel DBMSs}
\label{ch:RW0}

There have been some research efforts on
improving scalability of the parallel DBMS to the level of
NoSQL systems by carefully limiting certain functionalities
of the DBMS. Commercial parallel DBMSs for big data analytics such as
Teradata, GreenPlum, Sybase, and Vertica have shared-nothing MPP
architectures. Thus, these systems improve scalability by adding
multiple machines to the network
where dependencies among nodes are minimized. 
However, they do not take advantage of NoSQL system features such as
scalability, fault-tolerance, and load-balancing.

There are some research works that support limited DBMS
functionality optimized for handling a specific target workload in
the DBMS. PNUTS by Yahoo!\,\cite{CRS+08} supports scalability with eventual
consistency. Eventual consistency relaxes the
consistency level compared to strong consistency usually supported
by the DBMS. ODYS by KAIST\,\cite{Wha13} shows that a large-scale search engine
can be constructed by a shared-nothing massively-parallel
configuration of the DBMS using tight integration of DB and IR
features.

\vspace*{-0.3cm} % orphan

\subsection{NoSQL Systems}
\label{ch:RW1}

NoSQL systems are designed to easily add new machines
for scalability and to support load balancing among heterogeneous
machines. In addition, they support fault-tolerance by replicating
data into multiple nodes.

The components of typical NoSQL systems can be
classified into four layers with respect to their
functions\,\cite{Bud09, Wha13}: (1) the storage layer, (2)
key-value store layer, (3) parallel execution layer, and (4)
language layer. The storage layer provides the DFS. Representative
systems include GFS\,\cite{GGL03} and HDFS\,\cite{HDFS}.
The key-value store layer provides management
of data stored in the DFS in a key-value pair format.
Representative systems include BigTable\,\cite{CDG+06}
and Hbase\,\cite{Hbase}. The parallel execution layer
provides systems for processing queries in parallel. The
representative systems are MapReduce\,\cite{DG04} and its
open-source version Hadoop. It splits a job to multiple tasks,
assigns them to nodes, and processes them
in parallel. The language layer provides translation from a query
written in a higher-level language like SQL to a MapReduce job.
Representative systems include Pig\,\cite{ORS+08} and
Hive\,\cite{TSJ+09}. A distributed lock manager, which manages
shared resources in a distributed environment is also a primary
component of a NoSQL system although it cannot be classified
into the above four layers. Representative systems include
Chubby\,\cite{Bur06} and Zookeeper\,\cite{Zookeeper}.

\begin{figure}[htb]
 \centering
   \vspace*{0.2cm}
 \includegraphics[width=0.58\columnwidth]{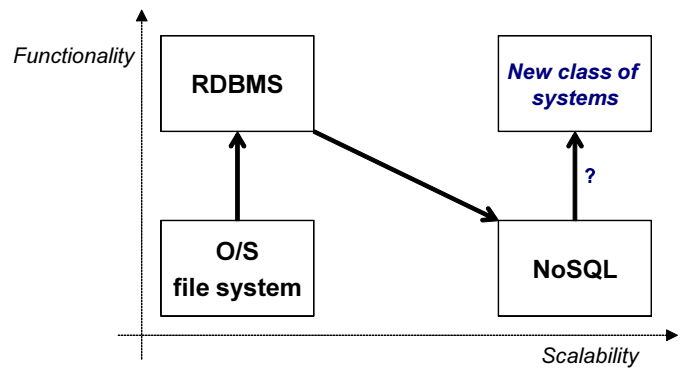}
 \vspace*{-0.5cm}
 \caption{Evolution of data management systems.}
  \vspace*{0.0cm}
 \label{fig:ch1}
\end{figure}

Figure \ref{fig:ch1} represents an evolutionary path of
data management systems. The RDBMS has been widely used due to its
rich functionality compared to an O/S file system, but it has
limitations in dealing with large-scale data. Therefore, to
overcome the limited scalability of the RDBMS, NoSQL systems
have been proposed emphasizing scalability and fault-tolerance.
The drawback of these systems, however, is that they lack the rich
functionality of the DBMS. Finally, the {\em new class of systems}
(the end of the arrow of Figure \ref{fig:ch1}), which supports
both scalability and DBMS-level functionality, has become an issue
in the literature\,\cite{Aba09}. Section \ref{ch:RW2} describes
these new systems. 

\subsection{Supporting DBMS Functionality on NoSQL Systems}
\label{ch:RW2}

In this paper, we classify efforts to support DBMS functionality
on NoSQL systems into two categories: (1) those on top of the
parallel execution layer\,\cite{MGL+10, ORS+08, TSJ+09} and (2)
those below the parallel execution layer\,\cite{ABA+09, Bra08, Dit10,
Dit12, Shu12}.

\subsubsection{Supporting DBMS Functionality on Top of the Parallel Execution Layer}

Most efforts have implemented a higher-level language layer on top
of the parallel execution layer\,\cite{Aba09}. Although MapReduce
allows us to easily execute parallel programs, programming in
MapReduce is still a burden. Thus, in order to provide a more
convenient programming environment, several methods have been
proposed for managing parallel tasks in MapReduce using a
higher-level language such as SQL. Specifically, these techniques
transform a user query into an equivalent MapReduce job and return
the results obtained by MapReduce to the user. The representative
systems are Hive\,\cite{TSJ+09} and Pig\,\cite{ORS+08}. They
belong to the language layer of Section \ref{ch:RW1}. Besides
high-level languages, methods of processing join operations using
MapReduce have also been studied\,\cite{BPE+10}.

\subsubsection{Supporting DBMS Functionality below the Parallel Execution Layer}

HadoopDB\,\cite{ABA+09} 
directly utilizes the RDBMS to process MapReduce tasks. HadoopDB
uses the RDBMS to process SQL queries for data analysis. That is,
when processing analytic tasks, HadoopDB first bulk-loads a
segment of the DFS data to the local database of each slave node
and uses the local DBMS of the node to process the data loaded in
the node. Parallel coordination is performed by MapReduce in the
master node. Since DBMS accesses the local data temporarily stored
in a node, HadoopDB suffers from duplicated storage. It also has
performance overheads due to loading DFS data to local databases.

Hadoop++ and HAIL support DBMS schemas and indexes in the
DFS using API functions provided by the DFS and MapReduce
library\,\cite{Dit10, Dit12}.
Especially, HAIL supports multiple clustering indexes by
clustering differently each replica of the DFS. However, they do
not support a higher-level language such as SQL.

Brantner et al.\,\cite{Bra08} has proposed a storage system
that supports transactions on top of Amazon's S3 system\footnote{a
distributed file system for Amazon cloud services}. It also
proposes methods of implementing various target consistencies in
Amazon S3. But, it lacks other DBMS functionalities such as the
query language and indexes.

The F1 DBMS\,\cite{Shu12} by Google supports some DBMS functionalities such
as SQL and transactions on top of the key-value store. Since F1
uses the key-value store, it can manage large-scale
data with scalability and fault-tolerance. The F1 DBMS is similar
to Odysseus/DFS proposed in this paper in that both systems have been designed
to support scalability and full features of a DBMS. F1 has some
DBMS functionalities such as multi-version transaction processing, secondary
indexes, and SQL. It also supports other features that satisfy
requirements of the AdWords business of Google.

However, F1\cite{Shu12} seems
to have the following limitations. First, F1 uses a key-value
store rather than a relational store. While a key-value store has the
disadvantage of redundantly representing the data, it
is not clear how F1 makes up for this disadvantage. 
Second, F1 proposes the concept of the hierarchical schema through
a series of one-to-many relationships. 
The hierarchical schema stores tuples that are pre-joined from
multiple relations in the physical storage. As a result, it cannot
support an efficient sequential scan of a relation for on the one-side of the
one-to-many relationship since tuples in the relation are
interleaved with those in their descendant relations. Furthermore, it is not clear whether it can also
efficiently support other complex schemas involving many-to-many
relationships.
These limitations are contrasted with Odysseus/DFS, which is
designed to integrate a full-blown RDBMS with the DFS. 

\section{Odysseus/DFS}
\label{ch:OdysseusDFS}

\subsection{The Architecture}
\label{ch:Concept}

\begin{figure}[htb]
 \vspace*{0.4cm}
 \centering
 \includegraphics[width=0.90\columnwidth]{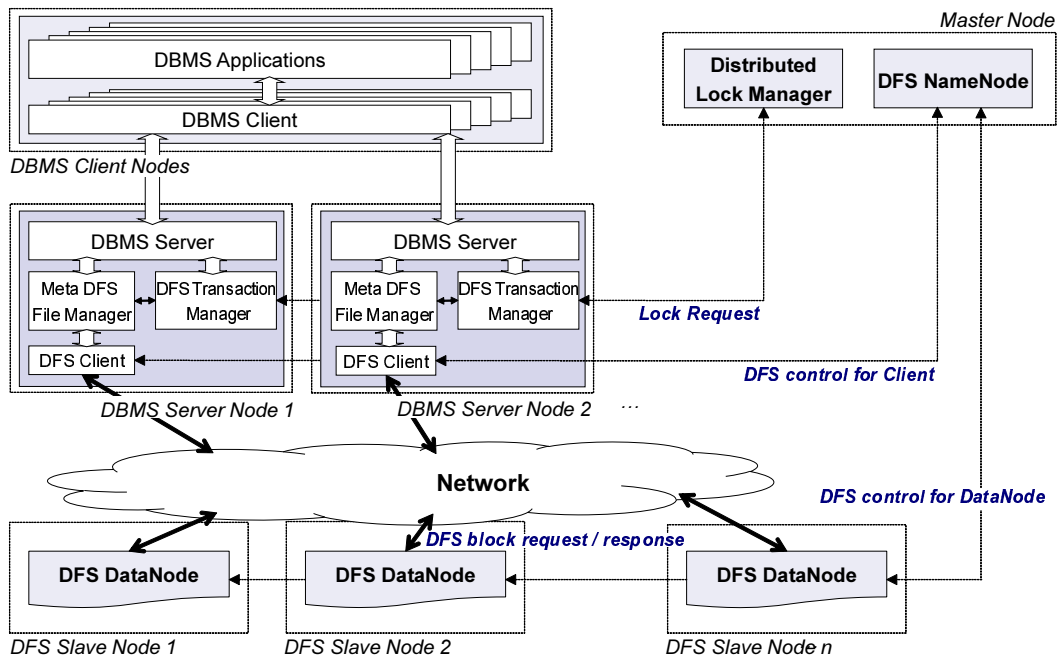}
 \vspace*{-0.5cm}
 \caption{The architecture of Odysseus/DFS.}
 \label{fig:arch}
  \vspace*{-0.1cm}
\end{figure}

Figure \ref{fig:arch} depicts the architecture of
Odysseus/DFS. Odysseus/DFS is composed of one master node,
multiple DBMS server nodes, and multiple DFS slave nodes. The master node
plays the roles of a DFS\footnote{Names of components of the DFS
are from those of HDFS\,\cite{HDFS}} NameNode and Distributed Lock
Manager. The DFS slave nodes contain DFS DataNodes. A DBMS server node
consists of a DBMS server, Meta DFS File Manager, DFS
Transaction Manager, and DFS Client.
A DBMS application residing in a DBMS client node accesses DBMS servers in the DBMS server nodes
through a DBMS Client.
The nodes described in the figure can be arbitrarily deployed.
For example, we can deploy DFS DataNode, DBMS client/server, and DBMS application in the same machine
or each of them in a different machine.
A DBMS server accesses data in the DFS
through the Meta DFS File Manager and manages transactions through the DFS
Transaction Manager.

The DFS NameNode manages the metadata of DFS files. A DFS DataNode
stores partitions of DFS files in duplicates. A DFS Client provides the user
interface necessary for using the DFS. The Distributed Lock Manager manages
locks for concurrency control.
A Meta DFS File Manager manages meta DFS files. A meta DFS file is a collection of DFS blocks
and supports in-place overwriting and appending
operations in the unit of a DFS block. The roles of the Meta DFS
File Manager are explained in Section \ref{ch:MDFmanager} in
detail. The DFS Transaction Manager performs data update,
recovery, and concurrency control so as to satisfy ACID properties
of the transactions in using the DFS as storage. They are elaborated
in Sections \ref{ch:Recovery} and \ref{ch:CC} in detail.

The architecture of Odysseus/DFS is novel and different from other existing
architectures proposed to manage large-scale data. The architectural difference
between Odysseus/DFS and other systems are summarized below:

\vspace*{-0.4cm}
\begin{itemize}
\item
Single node RDBMS: The RDBMS uses local disks, a disk array,
or a SAN as its storage while Odysseus/DFS uses the DFS.
Therefore, Odysseus/DFS takes advantage of the DFS such as
fault-tolerance, load balancing, and scalability while the RDBMS
cannot.

\item
Parallel DBMS (PDBMS): (1) Since, like in the RDBMS, the PDBMS stores
data in local disks, a disk array, or a SAN of slave nodes, it
cannot take advantage of the DFS.
(2) Since the PDBMS partitions the entire data into slaves
in a shared-nothing manner, each slave of the PDBMS cannot directly access
the data stored in the other nodes. Thus, the `explicit' communication among
multiple different DBMSs in the slaves is incurred. In contrast,
since a DBMS server of Odysseus/DFS accesses data through the DFS,
it can access any data stored in the other nodes
by the `implicit' communication conducted by the DFS,
providing the view of an integrated DBMS.
(3) The PDBMS does not know the place where the data necessary for processing a query reside.
Thus, to process a query, the PDBMS uses repartitioning or broadcasting, which incur an excessive communication cost
or uses semi-join like methods, which need pre-processing.
In contrast, a DBMS server of Odysseus/DFS can exactly locate data necessary for processing the query
through the DFS by using the physical page identifier provided by the DBMS---reducing the communication cost compared with the PDBMS.

\item
Key-value store:
(1) The key-value store provides only basic functions for data management
on the data represented in the key-value format while not providing full functionality of the DBMS.
(2) Since, like the PDBMS, the key-value store stores data in multiple slave nodes in a
shared-nothing manner, each slave node cannot directly access the data stored
in the other nodes.

\item
Language layer system: Systems  such as Pig or Hive
have a language layer, whose primary goal is to support a
high-level language for big data analytics on top of the MapReduce
framework. Since these systems primarily support queries and
workloads for analytics, they do not support other DBMS
functionalities (e.g., transaction processing).

\item
Google F1 DBMS: Google F1 DBMS introduces a
module for supporting SQL and transaction processing on top of a
key-value store (i.e., BigTable) based on the DFS. In contrast, in
Odysseus/DFS, the RDBMS is integrated directly on top of the DFS.
To implement full DBMS functionality, F1 should completely
re-implement a new query processor that can use a key-value store
as the storage. In contrast, the architecture of Odysseus/DFS is
RDBMS-independent. Hence, the Odysseus/DFS approach allows us to
use an existing RDBMS that already has full DBMS functionality by
adopting our Meta DFS File Manager and DFS Transaction Manager,
minimizing implementation overhead.

\end{itemize}

\vspace*{0.2cm} % orphan

\subsection{Characteristics of the DFS}
\label{ch:DFS}

The DFS is a system storing data by managing local storages of multiple
machines connected through a network. A DFS file is partitioned
and stored by the unit of the DFS block\footnote{In some documents, the term
`chunk' is used. In this paper, we use the term `DFS block' consistently.
The DFS block size is configurable (64MB by default).}. A DFS block is
replicated in multiple distinct DFS DataNodes for fault tolerance\footnote{The
number of replicas of a DFS block is also configurable (3 by default).}.
The DFS NameNode manages metadata describing the mapping between the
DFS file and DFS blocks that are stored in DataNodes and periodically checks health
of each DataNode to avoid data loss.

The DFS Client provides API functions for applications and
communicates with the NameNode and DataNodes. The DFS Client
supports the following four types of API functions: reading,
writing, renaming, and deleting a particular DFS file. The DFS has
the following characteristics in reading/writing data.

\vspace*{-0.4cm}
\begin{itemize}
\item
The DFS Client supports random reads in the unit of
the byte. Random reads in the DFS, however, are slower than
those in a local disk. While, in processing a random read, a local
disk incurs latency corresponding to the disk access time, the
DFS incurs additional overhead of
(1) metadata lookup from a DFS DataNode and
(2) network transfer when data are transmitted through the network.
We call this overhead the \emph{network transfer overhead}.
When we read from or write into a large-sized DFS
file, the network transfer overhead in the DFS is relatively
negligible since network transfer can be done in parallel with
disk read/write.

\item
When we sequentially read from or write into a large-sized DFS
file in another node, if
the network speed is slower than the disk
transfer rate, network speed becomes a bottleneck.
We call this overhead the \emph{network bottleneck overhead}.

\item
The DFS Client does not support API functions for in-place overwriting or appending
to an existing DFS file. Hence, the DFS is called a write-once-read-many storage.
If we have to overwrite in-place or append to the DFS file, we should perform
the following process that incur a very high overhead. First, we read the target
DFS file and store and modify it in a temporary storage. Next, we
delete the existing DFS file and write a new DFS file that has same file name
as the old one with the modified content. In this paper, we call this process a {\em DFS file remake}.

\end{itemize}

\subsection{Meta DFS File Manager}
\label{ch:MDFmanager}

We introduce an abstract data type, which we call {\em the meta
DFS file}, to manage the DFS file effectively.
For supporting updates to the database, the RDBMS must be able to
overwrite or append data to an
arbitrary position within the database.
However, the DFS does not support in-place overwrite and append
operations as described in Section \ref{ch:DFS}. The meta DFS file
supports not only read and write operations that are supported by
the DFS API, but also in-place overwrite and append operations.
To support these operations, the meta DFS
file partitions data into multiple DFS files, where each DFS file
consists of one DFS block. Thus, a meta DFS file is as an ordered
set of DFS files\footnote{The relationship between a meta DFS file
and its DFS files is analogous to that of an O/S file and its disk
blocks.}.

\ifeng Figure \ref{fig:meta} illustrates the architecture of the
Meta DFS File Manager that manages meta DFS files. The Meta DFS
File Manager accepts and processes a read, overwrite, and append
request to a particular meta DFS file from the DBMS server. A meta
DFS file maintains an ordered set of DFS files that belongs to it.
To maintain an ordered set, Meta DFS File Manager names each DFS
file as a concatenation of the meta DFS file name and {\em
block\_id} of the DFS file. For example, to store the meta DFS file
\verb"$(PATH)/data/" of size 256MB, Meta DFS File Manager manages
it as four DFS files of size 64MB and names each DFS file as
\verb"$(PATH)/data/00", \verb"$(PATH)/data/01",
\verb"$(PATH)/data/02", and \verb"$(PATH)/data/03", respectively.
The DFS NameNode maintains the \emph{Meta DFS File Table} for managing the DFS files that belong to each meta DFS file.
The attributes of Meta DFS File
Table are DFS file name, file size, the number of
blocks, the number of replicas, and the block position of each DFS
block. The block position stores as many DataNode IDs as the
number of replicas.

\begin{figure}[htb]
\vspace*{0.1cm}
 \centering
 \includegraphics[width=0.8\columnwidth]{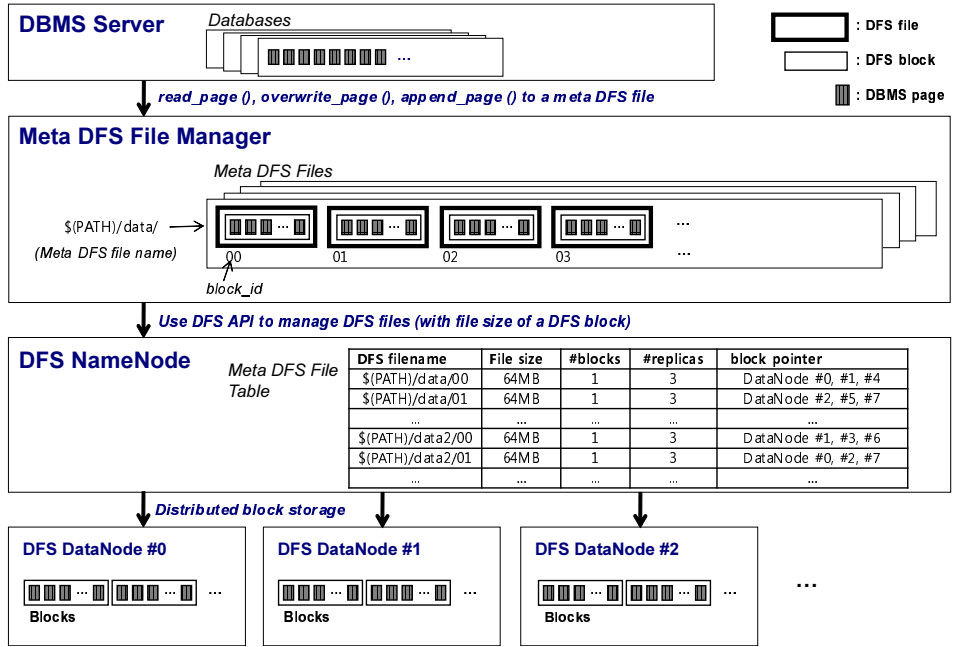}
 \vspace*{-0.5cm}
 \caption{The architecture of the Meta DFS File Manager.}
 \label{fig:meta}
\end{figure}

We also need to resolve syntactic differences between the DBMS
page and the DFS block. We
sequentially assign the \emph{pageid} to each page in a DFS block
to map the address space of the database to that of the meta DFS
file\footnote{The mapping is `static' since this mapping is between address spaces
and not between the data and address space. Thus, obviously, any update of data does not influence the mapping itself.}.
As a result, we can easily obtain the {\em block\_id} and {\em page\_offset} of a page
in a DFS block by a simple arithmetic operation. Specifically,
{\em block\_id} is ({\em pageid}/{\em N}) and {\em page\_offset}
is ({\em pageid} \texttt{mod} {\em N}), where {\em N} denotes the
number of pages in a DFS block.

\section{Recovery}
\label{ch:Recovery}

The recovery method we propose in this paper assumes that only one
DBMS transaction can update the database at a time. In other
words, the method assumes a serial schedule among write
transactions and a concurrent schedule among read transactions. In
the case where multiple processes reside in a single machine, we
can easily implement locks in the O/S shared memory so that we can
efficiently support locking. In the case where multiple processes
reside in multiple machines as in the DFS, however, there is no
efficient shared data structure such as the shared memory at the
O/S level.
Thus, we need an auxiliary system for synchronization among processes in multiple machines.
We use Zookeeper\,\cite{Zookeeper} for this
purpose. Zookeeper allows DBMS processes to synchronize with one
another by providing remote API functions to create, access, and
modify shared data structures. The details of implementation will
be presented in Section \ref{ch:CC}. Since the interprocess
communication through Zookeeper
is much slower than through O/S shared memory, the performance of
concurrency control could degrade severely. Therefore, using a large lock granularity
is preferred to reduce the locking overhead.
In this paper, we employ the database lock as the lock granularity.

Recovery methods based on immediate update such as ARIES\,\cite{Moh92}
incur high locking overhead while guaranteeing high
write-intensive concurrency.
On the other hand, recovery methods based on deferred-update or
shadow-pages
incur low locking overhead\,\cite{Wha12}.
Therefore, we employ a recovery method based on the latter as the recovery
method for Odysseus/DFS.

The {\em shadow-page deferred-update} (simply, SPDU)
method\,\cite{Wha12} combines the two methods to make up for the drawbacks
of each method. In Section \ref{ch:RecoverySPDU}, we
describe the SPDU recovery method in detail. In Section
\ref{ch:RecoverySPDUDFS}, we propose SPDU/DFS that modifies the SPDU method to resolve
inefficiency when we apply it to Odysseus/DFS.

\subsection{The SPDU Method}
\label{ch:RecoverySPDU}

The deferred-update method\,\cite{Elm10} defers updates to a data page
until commit time. Pages updated during a transaction are
temporarily stored in a deferred update file. At commit time,
updated pages are copied into the original page locations of the
database. This method has an advantage of maintaining clustering
of the data regardless of repetitive updates in the data since it
copies the updated pages back to the original page location of the
database. On the other hand, it has a disadvantage of having
difficulty in maintaining consistency in reading the database,
which can be resolved only by reading both the database and the
deferred update file during query processing.

The shadow-page method\,\cite{Gra81} maintains the mapping between the logical and physical
pageid's using a page mapping table. Any page updated during a transaction is
allocated a new page in the database. The method maintains both the new
mapping table that reflects updates and the old one that does not.
The method simply selects the new mapping table at commit but the
old one at abort. The method does not incur the data inconsistency
problem of the deferred-update method. On the other hand, when a page is
updated, the method stores the updated page into a place different from the
original one. Hence, repetitive updates in the data cause
the clustering of the data destroyed.

The Shadow-Page Deferred-Update (SPDU) method\footnote{This
has been patented in the U.S.\,\cite{Wha12}.} \,\cite{Wha12} hybridizes the
deferred-update and the shadow-page methods. Hence, it has
advantages of both methods. The SPDU method works correctly for a
serial schedule with respect to write transactions and a
concurrent schedule with respect to read transactions.
Specifically, we store a database in a \emph{data file} and
store updated pages of the database in a \emph{log file},
which is separate from the data file.
Since a transaction deals with a database, a DBMS process for each
transaction manages a data file and the corresponding log file.
Whenever a DBMS process commits a transaction, the log file
is reflected into the corresponding data file, and
then, the log file is initialized. Thus, the log file allows to defer
updating the data file, which plays the role of a deferred update
file in the deferred-update method.

The deferred update method has the overhead of accessing both the
data file and deferred update file to read a page. The SPDU method
removes this overhead by using an in-memory data structure called
the \emph{log table index}. The log table index stores the offsets
of the pages stored in the log file. Thus, we can find out whether
a certain page is stored in the log file or in the data file by
searching the log table index.
The log table index in SPDU is similar to the page mapping table
in the shadow-page method in the sense that they refer to the most
recent position where an updated page is stored. But, the difference
is that all the updates in the log table index are reflected back to the data
file when the transaction commits initializing the log table index while, 
in the shadow-page method, they are retained in the newly allocated pages
pointed to by the updated page mapping table.

An entry of the log table index consists of $<${\em pageid}, {\em
offset}$>$ pair. Here, {\em pageid} is the pageid of an original
page in the data file, and {\em offset} is the offset for the page
in the log file (i.e., a physical pageid in the log file). The
entire set of pages of the database is stored in the data file. If
we update a page in the data file, we append the updated page in
the log file. We also store $<${\em pageid}, {\em offset}$>$ pair
for the page into the log table index. Figure
\ref{fig:logtableindex} shows page mapping in the SPDU method
using the log table index. In the data file, {\em pageid} of a
page is equal to {\em offset} of the page since the DBMS maintains
the pages in the order of $pageid$. However, in the log file, {\em
pageid} of a page is different from {\em offset} of the page since
we store updated pages according to the order updates
occurred. In Figure \ref{fig:logtableindex}, we update pages 3, 7,
1, and 9 in this order. When we need to access a page, we need to
find out whether the page is stored in the data file or in the log
file by searching the log table index. If it is
stored in the log table index (e.g., pages 1, 3, 7 or 9 in
Figure~\ref{fig:logtableindex}), we access the log file by using
the {\em offset} stored in the log table index. Otherwise, we
access the data file by using $pageid$ of the page as the offset
itself.

\begin{figure}[!ht]
\vspace*{0.2cm}
 \centering
 \includegraphics[width=0.60\columnwidth]{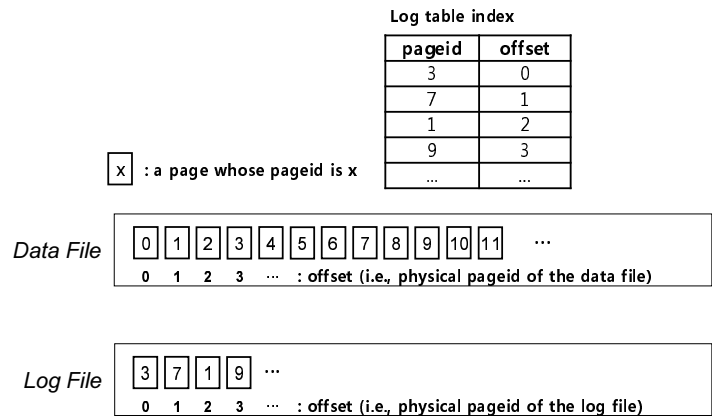}
 \vspace*{-0.6cm}
 \caption{Page mapping using the log table index in SPDU method.}
  \vspace*{-0.0cm}
 \label{fig:logtableindex}
\end{figure}

Figure \ref{fig:spdu} shows the algorithms for six basic
operations of the SPDU method: (1) writing a page, (2) reading a
page, (3) committing a transaction, (4) aborting a transaction,
and (5) restarting the system, and (6) post commit processing for
transaction commit and system restart.

\begin{figure}[!ht]
\vspace*{-0.8cm}
 \centering
 \includegraphics[width=0.85\columnwidth]{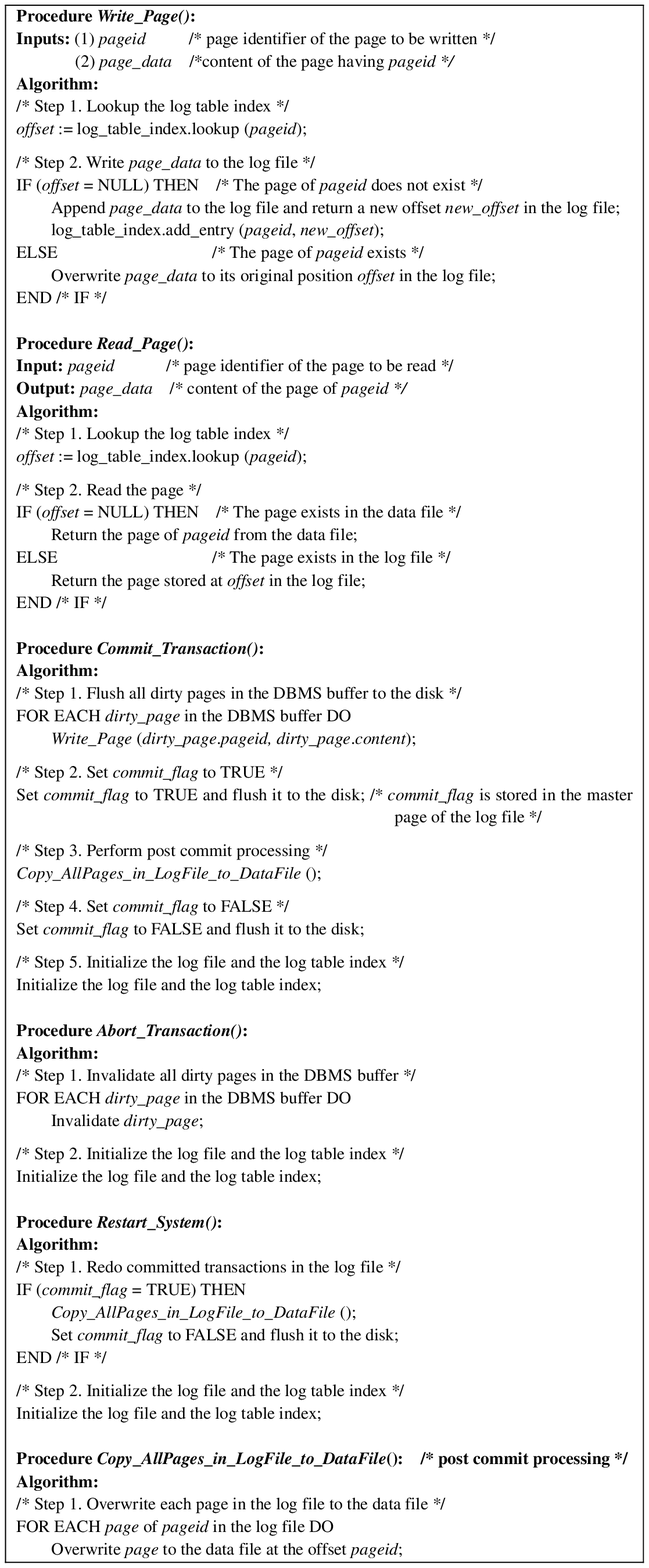}
 \vspace*{-1.5cm}
 \caption{Algorithms for the Shadow-Page Deferred-Update (SPDU) method.}
  \vspace*{-1.2cm}
 \label{fig:spdu}
\end{figure}

\emph{Write\_Page} takes
{\em pageid}, the page identifier, and
{\em page\_data}, the content of the page, as inputs.
In Step 1, we look up the log table index to find whether the page to be written
exists in the log file.
In Step 2, if it exists in the log file,
we overwrite it with the {\em page\_data};
if not, we allocate a new page in the log file,
write {\em page\_data} into the new page, and add an entry $<${\em pageid}, {\em offset}$>$ to the log table index.

\emph{Read\_Page} takes
{\em pageid} as the input
and returns {\em page\_data}, as the output.
In Step 1, we look up the log table index to check whether the page to be read
exists in the log file.
In Step 2, if it exists in the log file,
we read the page from the log file using {\em offset};
if not, we read the page from the data file.

\emph{Commit\_Transaction} commits the transaction and
copies the pages updated by the transaction from the log file to
the data file. In Step 1, we flush every dirty page in the DBMS
buffer to the log file and to the disk. In Step 2, we set the {\em
commit\_flag} to \texttt{TRUE} in the master page\footnote{This is
the first page of the log file.} of the log file and flush it to
disk. The transaction commit is completed atomically during Step
2. In Step 3, we perform post commit processing by calling
\emph{Copy\_AllPages\_in\_LogFile\_to\_DataFile} procedure. The
procedure reads each page in the log file, gets its {\em pageid}
from the log table index, and copy its {\em content} into the
original place in the data file using {\em pageid}. In Step 4, we
set the {\em commit\_flag} to \texttt{FALSE} and flush it to disk
to mark that post commit processing has been completed. In Step 5,
we initialize the log file and the log table index.

\emph{Abort\_Transaction} rolls back every write performed
in the transaction. In Step 1, we invalidate every dirty page in
the DBMS buffer by setting valid flag for the page to
\texttt{FALSE}. In Step 2, we initialize the log file and the log
table index.

\emph{Restart\_System} initializes the system. In Step 1,
if the system has crashed during the post commit processing (i.e.,
if {\em commit\_flag} is \texttt{TRUE}), we restart the post
commit processing\footnote{When crash occurs during post commit
processing, consistency of the data file is violated. To recover
consistency, post commit processing must be repeated from the
beginning to the end. Thus, post commit processing is made
idempotent\,\cite{Wha12} to make repetitive crashes and restarts
of post commit processing still guarantee consistency of the data
file if completed to the end.}. In Step 2, we initialize the log
file and the log table index to roll back the updates done in the
uncommitted pages. 
\subsection{Shadow-Page Deferred-Update Recovery Method for DFS (SPDU/DFS)}
\label{ch:RecoverySPDUDFS}

In this section, we present the \emph{shadow-page deferred-update recovery method for DFS}
(simply, SPDU/DFS)
that modifies the SPDU method to use the DFS as the storage. The
modifications for SPDU/DFS are as follows: (1) We manage the
storage in the unit of the meta DFS file instead of the file.
(2) We propose enhanced techniques to resolve inefficiencies of
the SPDU algorithms incurred by the characteristics of the DFS.

By managing the data and log file by the Meta DFS File
Manager, the SPDU/DFS method has the following strengths.
(1) If the data file were composed of only one DFS file, a DFS file remake
would occur for the entire data file whenever the system copies updated
pages in the log file to the data file during a transaction
commit. In the SPDU/DFS method, by managing a data file as a meta
DFS file, the system performs DFS file remake only in the unit
of the DFS block. (2) Likewise, if the log file were composed of one DFS file, a
DFS file remake would occur for the entire log file whenever the system
appends a page to the log file.
By using a meta DFS file for the log file, however, we can have the same
effect as with the data file.

The rest of the section describes further enhancements to
solve inefficiencies that would arise when we directly apply the SPDU
method to the DFS.
Since the DFS block, a unit of I/O for a meta DFS file, is much larger than the DBMS page,
inefficiency is incurred when we need to modify only a tiny part of the DFS block.
In Section \ref{ch:spdudfs1}$\sim$\ref{ch:spdudfs3}, we present three techniques for enhancement
to alleviate this problem.

\subsubsection{DFS block update buffer}
\label{ch:spdudfs1}

To reflect the updated pages into the log meta DFS file in the
unit of the DFS block, we introduce a {\em DFS block update
buffer} whose size is identical to that of one DFS block. First,
we accumulate the updated pages in the DFS block update buffer.
This is possible since updated pages are written sequentially
in the log file.
Then, the system writes out the buffer when it is full or at transaction
commit. This reduces the number of DFS file remakes.

Due to the DFS block update buffer, we should modify the page mapping
of SPDU for SPDU/DFS. Figure \ref{fig:logtableindex2} shows the modified page
mapping.
When we access a page, we should consider not only the data
and log files but also the DFS block update buffer since
recently updated pages reside in the buffer. Thus, when we
access a page, we first check whether the page resides in the DFS
block update buffer. If it does, we access it from the DFS
block update buffer; otherwise, we choose the data meta DFS file
or log meta DFS file using the log table index in the same way as
in SPDU. In the log table index, we modify the offset in a log table entry to
a pair $<${\em block\_id}, {\em b\_offset}$>$ where {\em b\_offset}
is an offset within the DFS block having {\em block\_id}.

\begin{figure}[!ht]
\vspace*{0.2cm}
 \centering
 \includegraphics[width=0.72\columnwidth]{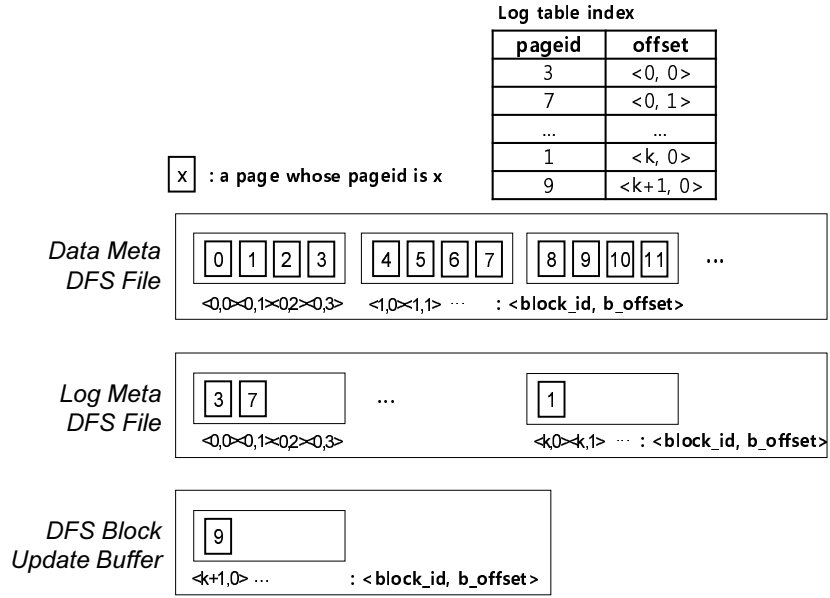}
 \vspace*{-0.6cm}
 \caption{Page mapping in SPDU/DFS.}
  \vspace*{-0.0cm}
 \label{fig:logtableindex2}
\end{figure}

\subsubsection{Post Commit Processing in the Unit of the DFS block }
\label{ch:spdudfs2}

In SPDU, during \emph{Commit\_Transaction} and
\emph{Restart\_System}, we copy updated pages in the log file to
the data file (i.e., post commit processing). This post commit
processing process would become extremely inefficient for SPDU/DFS
since a DFS file remake occurs every time a page is copied. This
inefficiency stems from the fact that updates pages for a DFS
block in the data meta DFS file are stored scattered in the log meta DFS
file. To resolve this problem, we sort the pages in the log meta DFS file to be able to copy the updated pages
to the data meta DFS file in the unit of the
DFS block. Specifically, (1) we sort all the updated pages in the
log meta DFS file according to {\em pageid} and group them in the
unit of the DFS block. (2) We load each DFS block containing the
updated pages from the data meta DFS file to the main memory,
update the DFS block with the updated pages in the log
meta DFS file, and write it back to the data meta DFS file through a DFS file remake.
This incurs only one DFS file remake for one DFS block instead
of for one page. In addition, when there have been multiple updates in
the same page,
this method allows us to reflect it only once by using the last update to the page.

\subsubsection{Deferred Post Commit Processing}
\label{ch:spdudfs3}

In Section \ref{ch:spdudfs2}, we improved the efficiency of
post commit processing. But, post commit processing is inherently
inefficient since the DFS file remake is an expensive operation
itself\footnote{In the worst case, each page of a DFS block in the
log meta DFS file can be stored in a different DFS block in the data
meta DFS file. This indicates that $\frac{64MB}{4KB}=16,000$ DFS file
remakes could occur to copy one DFS block in the log meta DFS file
back to the data meta DFS file, if the size of a DFS block is 64MB and that of a page is 4KB.}.
To alleviate this problem, we defer post commit processing until
multiple transactions commit rather than doing it after each
commit.
Specifically, we perform post
commit processing in a batch
(1) periodically or (2) when the size of
the log meta DFS file exceeds a pre-determined value.

Due to deferred post commit processing, the log meta DFS file
normally contains not only uncommitted data but also committed
data. Thus, two new issues arise. First, we need
to distinguish committed data from uncommitted data in the log meta
DFS file to correctly process transaction abort or
system restart.
For this, we introduce a new flag in each DFS block in the log meta DFS file called {\em
commit\_complete}. If {\em commit\_complete} is
\texttt{TRUE}, the data in the DFS block must have been committed; if not,
the data have not been committed. The {\em commit\_complete} flag plays the role of
a checkpoint allowing us to know that the data in the DFS block with {\em
commit\_complete} set \texttt{TRUE} and in all the prior DFS blocks
have been committed since we only deal with serial schedules of write operations.
Therefore, when the transaction
is aborted or the system is restarted, we delete DFS blocks in the
log meta DFS file from the most recent DFS block in the reverse order
until we meet a DFS block with the {\em commit\_complete} flag set
\texttt{TRUE}.

Second, when multiple transactions are working simultaneously, a
transaction may need to access updated data committed by
another transaction. To access the most recent data, each DBMS
process should keep the log table index up to date reflecting the most
recent state of the database. For this, we reconstruct the log
table index for a DBMS process by reading the log meta DFS file at
the time when the process acquires a read or write lock\footnote{For efficient
reconstructing of the log table index, we store a list of pages
contained in each DFS block at the end of the DFS block. We can
reconstruct the log table index by accessing only the list of each
DFS block without having to read the entire set of pages in the
DFS block.}. Since we use the database lock, the database cannot be
updated by other DBMS processes after acquiring the lock.
Therefore, it guarantees that each DBMS process can access the
most recent state of the database. Figure \ref{fig:spdudfs2} shows
an example of reconstructing the log table index when two DBMS
processes (P1 and P2) are running simultaneously. Initially, there
is no updated pages in the database, and then, P1 acquires a lock,
updates pages 5 and 3, and releases a lock. Thus, these updates
are reflected to the log meta DFS file and the log table index for
P1. When P2 acquires the lock, we reconstruct the log table
index for P2 to reflect the most recent state of the database,
i.e., to include pages updated by P1.

\begin{figure}[!ht]
\vspace*{0.2cm}
 \centering
 \includegraphics[width=0.85\columnwidth]{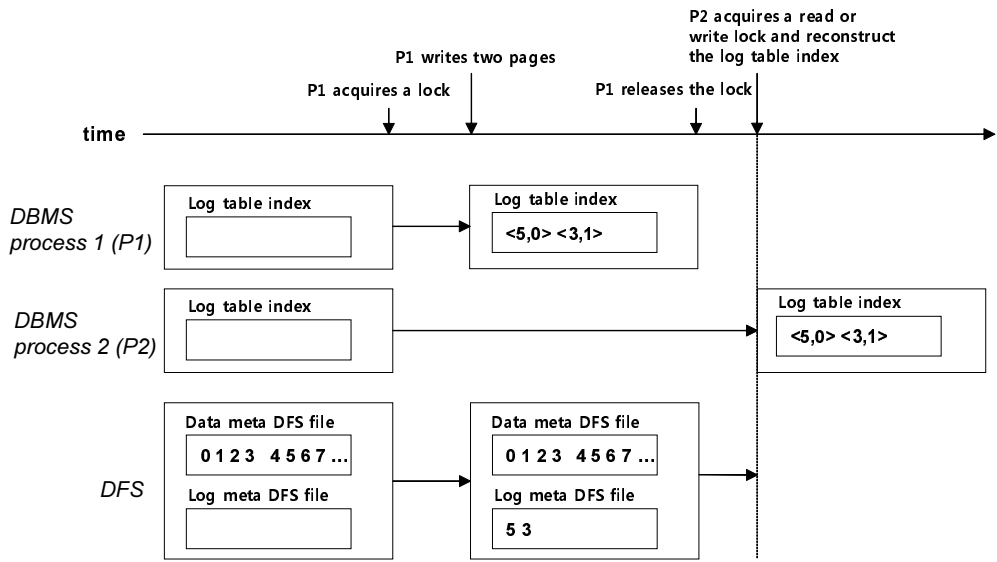}
 \vspace*{-0.6cm}
 \caption{Reconstruction of the log table index.}
 \label{fig:spdudfs2}
\end{figure}

\section{Concurrency Control}
\label{ch:CC}

In this section, we present the concurrency control method of
Odysseus/DFS. Odysseus/DFS uses a locking-based concurrency
control method.
We present the way we implement read/write lock
operations for concurrency control.

An important property of a lock is its granularity. As
discussed in Section \ref{ch:Recovery}, we use the database lock for
concurrency control to achieve serial schedules for write transactions.
We implement read/write locks using Zookeeper\,\cite{Zookeeper} as
follows.
A DBMS process that wants to acquire a lock connects to Zookeeper,
requests the lock, and wait until acquiring the lock. When the
process satisfies conditions to acquire the lock, it finishes
waiting and acquires the lock. We also implement a lock
compatibility matrix realizing write/write and read/write conflicts
using Zookeeper as follows. A process that
requested a read lock acquires it after all the write locks
requested prior to the read lock have been released. A process
that requested a write lock acquires it after all the (read or
write) locks requested prior to the write lock have been released.

A detailed implementation is as follows.
Zookeeper manages a data structure similar to an O/S directory/file. We implement locks
by using the API functions for managing Zookeeper files since
Zookeeper itself does not support API functions controlling locks
directly. We create a Zookeeper directory for a data meta DFS
file and create a Zookeeper file in the directory for a particular
DBMS lock request.
Specifically, when a DBMS process wants to acquire
a lock for a data meta DFS file, it creates a Zookeeper file
with the file name of the form ($lock\_type$, $lockid$)
in the Zookeeper
directory representing the data meta DFS file.
Here, \emph{lock\_type} is either a read lock or a write lock;
\emph{lockid} is incrementally assigned by
Zookeeper in the order of the time of lock request. When a lock creation fails or
a lock is released, the corresponding Zookeeper file is destroyed.

Locks are granted in the order of lockid¡¯s, i.e., in the order the lock requests arrive at Zookeeper.
Specifically, when a DBMS process requests a read lock, if there
are only read locks or there are write locks that have a larger
lockid than that of the current lock request,
it instantly acquires the
read lock. However, if there are write locks that have
lockid's smaller than that of the read lock requested, it sets a {\em
watcher} to the write lock that has been requested last and waits.
When the lock observed by the watcher is released, the current process
is awakened and acquires the read lock.

When a DBMS process requests a write lock, it instantly acquires
the write lock only when all the lockids of the existing locks, except
a read lock that is possibility owned already by the DBMS process itself, are bigger than the lockid
of the write lock requested. However, if there is any lock that has
a lockid smaller than the write lock requested, the current process waits after setting
a watcher to the lock that has been requested last.
When a lock observed by the watcher
is released, the current process is awakened and acquires the write lock.

\vspace*{0.4cm} % orphan

\section{Performance Evaluation}
\label{ch:Performance}

\subsection{Experiment Setting}
\label{sec:ExprSetting}

In this section, we compare the performance of Odysseus/DFS
with that of Hbase, a representative NoSQL system, to show the
effect of using the DBMS functionality.
In particular, we show the
effect of using the index since it affects the
performance of the system. The other DBMS functionalities make
Odysseus/DFS more usable than Hbase
but does not much affect the performance.
We also compare the performance of Odysseus/DFS with that of an
RDBMS using local storage to show that the performance
overhead of Odysseus/DFS is not significant despite
supporting scalability and reliability through the DFS, which are not supported by
the RDBMS.
We conduct experiments on read and write operations for sequential
and random workloads: (1) \emph{sequential read},
(2) \emph{sequential write}, (3) \emph{random read}, and
(4) \emph{random write}.

In order to set up Odysseus/DFS and Hbase,
we use a cluster of nine nodes: one master and eight slaves.
Each node consists of 3.2GHz Intel Quad-Core CPU, 8GB RAM and one 1TB hard disk.
Nodes are connected by a 1Gbps network switch. The average transfer rate of
the hard disk is 120MByte/s. The average network transfer rate is 80MByte/s.

To implement Odysseus/DFS, we use the Odysseus DBMS\,\cite{Wha05, Wha14} (coarse-granule
locking version),
replacing the file system its own and transaction manager with the Meta DFS File Manager and the DFS Transaction
Manager. We also use the Odysseus DBMS when we conduct experiments for
the RDBMS using the local storage for a fair comparison. We use
the entire database as the locking granule of the Odysseus/DFS and
the RDBMS\,(Odysseus) to employ a large locking granule
realizing serial schedules for write transactions\footnote{Odysseus
supports not only large locking granularity, i.e., database-level
locking, but also small locking granularity, i.e., record-level
locking.}.
As the storage for Odysseus/DFS and
Hbase, we use HDFS (binary version
1.0.3) from \url{http://hadoop.apache.org}. We use Hbase binary
version 0.94.7 with default settings for configuration. Odysseus DBMS has been implemented in 450,000
lines of precision C and C++ codes\,\cite{Wha05, Wha13, Wha14}.
Meta DFS File Manager of Odysseus/DFS is
implemented in C on top of the C language API of DFS Client.
HDFS and Hbase are written in Java.

In the experiments, we use the synthetic data set generated by Pavlo
et al.\,\cite{Pav09}\footnote{This data set was also used by several other
studies\,\cite{ABA+09, Dit10, Dit12} on big data management.}.
This data set models rankings and visit logs of web pages. It has three tables:
Web Document, Ranking, and UserVisits. For our experiments,
we use the UserVisits table, which has the largest number of tuples among the three tables.
There are 155 million tuples in the UserVisits table. The schema of the database
is described in Figure \ref{fig:Schemas}. For the experiments, we choose
the visitDate attribute to cluster the UserVisits table. Since Hbase does
not support SQL, we hand-coded the schema equivalent to the one defined
in SQL. Specifically, we map an attribute of SQL to a column of Hbase.

\begin{figure}[!ht]
\begin{center}
\begin{boxedminipage}[c]{16.7cm}
\vspace*{-0.2cm}

\vspace*{0.3cm} \noindent\verb"(SQL)"

\vspace*{-0.3cm} \noindent\verb"CREATE TABLE UserVisits ( sourceIP VARCHAR(16), destURL VARCHAR(100), visitDate DATE,"

\vspace*{-0.3cm} \verb"                        adRevenue FLOAT, userAgent VARCHAR(64), countryCode VARCHAR(3),"

\vspace*{-0.3cm} \verb"                        languageCode VARCHAR(6), searchWord VARCHAR(32), duration INT );"

\vspace*{-0.15cm} \noindent\verb"CREATE INDEX uservisits_sourceip_idx ON UserVisits ( sourceIP );"

\vspace*{0.2cm}

\vspace*{0.3cm} \noindent\verb"(HBase)"

\vspace*{-0.3cm} \noindent\verb"HBaseAdmin hbase = new HBaseAdmin(config);"

\vspace*{-0.3cm} \noindent\verb"HTableDescriptor tdesc = new HTableDescriptor(tableName);"

\vspace*{-0.3cm} \noindent\verb"String[] colName = new String[] {`sourceIP', `destURL', `visitDate', `adRevenue',"

\vspace*{-0.3cm} \noindent\verb"                   `userAgent', `countryCode', `languageCode', `searchWord', `duration'};"

\vspace*{-0.3cm} \noindent\verb"for(int i=0; i<colName.length; i++) {"

\vspace*{-0.3cm} \noindent\verb"    HColumnDescriptor cdesc = new HColumnDescriptor(colName[i].getBytes());"

\vspace*{-0.3cm} \noindent\verb"    tdesc.addFamily(cdesc);"

\vspace*{-0.3cm} \noindent\verb"}"

\vspace*{-0.3cm} \noindent\verb"hbase.createTable(tdesc);"

\end{boxedminipage}
\caption{The schema of the database used in the experiments.}
\label{fig:Schemas}
\end{center}
\vspace*{-0.4cm}
\end{figure}

Figure \ref{fig:Queries} shows the queries used for the experiments.
Odysseus/DFS and the RDBMS using local storage process the queries through SQL, and Hbase
through a hand-coded Java program equivalent to SQL.
The \emph{Scan} query reads 100,000 tuples of the table
sequentially.
The \emph{Insert} query inserts 10,000 new tuples to the table
sequentially.
To make the sequential write workload for the
Insert query, we inserted the data to be ordered by the clustering attribute.
The \emph{Select} query retrieves tuples satisfying the
condition where the sourceIP attribute is \verb"160.110.44.44" in the
table\footnote{The number of tuples that satisfy the condition is 70; thus, the
selectivity is $4.5\times10^{-7}$.}. This query reads data randomly when the
non-clustering attribute sourceIP has a secondary index. Otherwise, it scans
the entire table sequentially.
The \emph{Update} query finds the tuples satisfying the condition where
the sourceIP attribute is \verb"160.110.44.44" and updates the countryCode
attribute to \verb"`ABC'".
As the Select query, it reads data randomly only if the sourceIP has a secondary index.
%
%
%To make the random read/write workload for Select/Update query, we use the
%queries that have a condition on the non-clustering attribute sourceIP having a secondary index defined on it.

\begin{figure}[!ht]
\begin{center}
\begin{boxedminipage}[c]{15.7cm}

\vspace*{0.3cm}   \noindent Scan : \verb"SELECT * FROM UserVisits LIMIT 100000;"

\vspace*{-0.15cm} \noindent Insert : \verb"INSERT INTO UserVisits VALUES ( ... ); (repeat 10000 times)"

\vspace*{-0.15cm} \noindent Select : \verb"SELECT * FROM UserVisits WHERE sourceIP = `160.110.44.44';"

\vspace*{-0.15cm} \noindent Update : \verb"UPDATE FROM UserVisits SET countryCode = `ABC'"

\vspace*{-0.3cm} \verb"            WHERE sourceIP = `160.110.44.44';"

\vspace*{0.2cm}
\end{boxedminipage}
\caption{The SQL queries used in the experiments.}
\label{fig:Queries}
\end{center}
\vspace*{-0.3cm}
\end{figure}

The elapsed time of the SQL queries is measured.
All the experiments are performed in cold start; in order to obtain
consistent results, we flush the DBMS buffers, O/S file buffers, and disk
buffers before executing each query. We average the elapsed times of
five identical executions of each query.

\subsection{Performance Results}
\label{sec:PerformanceResults}

\subsubsection{Comparison with NoSQL}

In this section, we compare performance of
Odysseus/DFS and Hbase. While Hbase
cannot support a secondary index, Odysseus/DFS does. Thus,
Odysseus/DFS outperforms Hbase for the queries that include
predicates for the non-clustering attribute having a secondary index. Among the
queries in Figure \ref{fig:Queries}, Scan and Update are such queries.

Figure \ref{fig:new1} shows the performance of
Odysseus/DFS and Hbase for Scan and Update queries.
Odysseus/DFS (w/w index) represents the performance of
Odyssesus/DFS using a secondary index on the sourceIP attribute.
Odysseus/DFS (w/o index) represents the performance without
using an index. The result shows that Odysseus/DFS
(w/o index) is faster than Hbase by 2.2 times, and Odysseus/DFS
(w/w index) is faster than Hbase by 18.6$\sim$27.2 times. First,
let us focus on the former result. It
is difficult to directly compare Odysseus/DFS with Hbase since the former
uses a DBMS while the latter a key-value store. However, we conjecture that
the primary reasons for the performance difference are from the storage structures and the
programming languages. Specifically, Hbase stores the data in
column-store. Thus, processing queries involving multiple
attributes is inefficient. Moreover, Hbase is implemented in Java
while Odysseus/DFS in C. In this paper, we focus on the
performance difference by the DBMS functionality, i.e., secondary
indexes, rather than the system-oriented differences. Second, let us focus on the latter result.
Here, if we remove the system-oriented performance difference (i.e., 2.2 times),
we observe that the actual performance
enhancement due to using indexes in Odysseus/DFS compared with Hbase is 8.6$\sim$12
times.

\begin{figure}[!ht]
\begin{center}
 \vspace*{-0.2cm}
 \includegraphics[width=0.63\columnwidth]{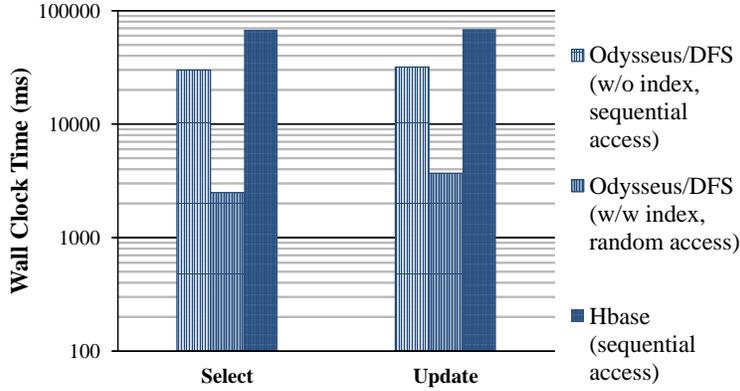}
 \vspace*{-0.7cm}
\caption{Results of Scan and Update queries of Odysseus/DFS and Hbase.}
 \vspace*{-0.0cm}
\label{fig:new1}
\end{center}
\end{figure}

\subsubsection{Comparison with an RDBMS using local storage}

In this section, we show performance overhead incurred when the DBMS uses
the DFS in place of local disk as the storage. We conduct experiments
for four workloads discussed above.
Figure \ref{fig:new2} shows the result of sequential read and sequential write obtained by running Scan and Insert queries.
In sequential read, Odysseus/DFS has 31\% additional overhead compared to Odysseus
with local storage. This overhead is due to the network bottleneck overhead
described in Section \ref{ch:DFS}, i.e., it is incurred
when the network speed cannot catch up with the transfer rate of magnetic disks.
In sequential write, the performance of Odysseus/DFS is 19\% faster than that of Odysseus
with local storage
since the former uses deferred post commit processing.
Since post commit processing can be done
in the background when the workload is not heavy,
its overhead can be saved.

\begin{figure}[!ht]
\begin{center}
 \vspace*{0.2cm}
 \includegraphics[width=0.65\columnwidth]{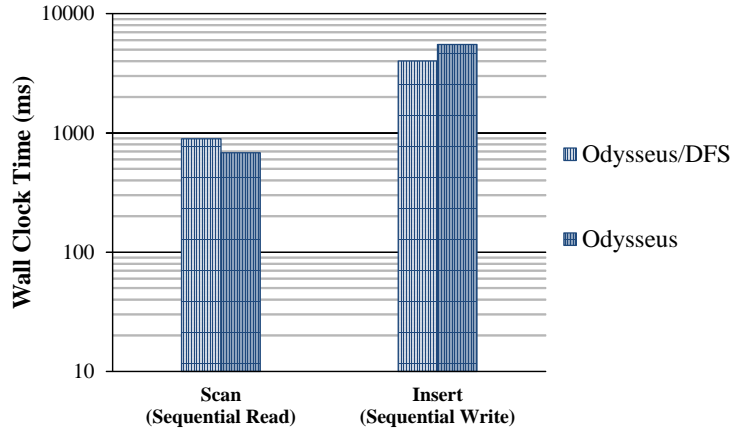}
 \vspace*{-0.7cm}
\caption{Results of sequential read and write operations of Odysseus/DFS and Odysseus.}
 \vspace*{-0.2cm}
\label{fig:new2}
\end{center}
\end{figure}

Figure \ref{fig:new3} shows the results of random
read and random write obtained by running Select and Update queries. In random read, Odysseus/DFS
has 67\% additional overhead compared to Odysseus. This overhead
is due to the network transfer overhead described in Section
\ref{ch:DFS}, i.e., it is incurred by network transfer time
for the data pages randomly accessed through the DFS.
In random write, Odysseus/DFS has 54\% additional overhead compared to Odysseus
due to a similar reason. To
process the random write query, the system first searches
for the tuples that satisfy the condition and then update them.
Most of the query processing time is for searching since only 70
tuples out of 155 million tuples read are updated, rendering the
updating time negligible. Therefore, the result of random
write has a similar tendency to that of random
read.

\begin{figure}[!ht]
\begin{center}
 \vspace*{0.3cm}
 \includegraphics[width=0.65\columnwidth]{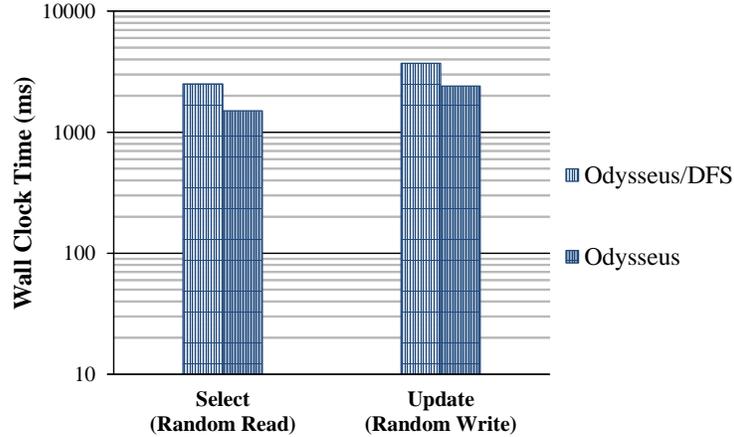}
 \vspace*{-0.7cm}
\caption{Results of random read and write operations of Odysseus/DFS and Odysseus.}
 \vspace*{-0.2cm}
\label{fig:new3}
\end{center}
\end{figure}

We observe the following from the experiments:

\vspace*{-0.4cm}
\begin{itemize}
\item
The performance of Odysseus/DFS is comparable to that of Odysseus even though
the former supports scalability by using the DFS as the storage.
It is shown to be 0.73 $\sim$ 1.67 times that of Odysseus.
Odysseus/DFS has performance improvement over Odysseus in sequential write
since it defers post commit processing as discussed in Section \ref{ch:spdudfs3}.
It has performance degradation compared to Odysseus
in sequential read since the network bandwidth does not catch up with the disk
transfer rate, and in random read and random write, due to network transfer time.
These are unavoidable in a distributed environment.

\end{itemize}

\vspace*{-0.4cm} % orphan

\section{Conclusions}
\label{ch:Conclusions}

\vspace*{-0.2cm} % orphan

The contributions of the paper are as follows.
First, we have proposed a new architecture integrating
an RDBMS with the DFS for transaction processing of big
data. By using the DFS as the storage of an RDBMS, we
can extend the scalability and reliability of the RDBMS. Second,
we have proposed the notion of the meta DFS file
to be used as the storage for an RDBMS allowing
in-place overwriting and appending operations to the
DFS by the unit of the DFS block.
Third, we have presented techniques for transaction management that support
update, recovery, and concurrency control while considering the
characteristics of the DFS. The transaction method we implemented in
this paper supports only serial schedules with respect to write
transactions and concurrent schedules with respect to read
transactions employing coarse-granularity locking in order to reduce the
locking overhead. Fourth, we have presented experimental results to
compare the performance of the RDBMS using the DFS as the storage
with those of NoSQL-based key-value store and the RDBMS using local
disk as the storage.
The result shows that Odysseus/DFS has an advantage over Hbase\,(18.6$\sim$27.2 times)
due to availability of secondary indexes
and over relational DBMS in sequential write due to deferred post commit processing;
it has a disadvantage over a relational DBMS in sequential read\,(31\%) due to the network bottleneck overhead
and random read/write\,(54\%$\sim$67\%) due to the network transfer overhead.

The strengths of Odysseus/DFS are as follows. First, it has DBMS-level
functionality compared to NoSQL systems.
That is, Odysseus/DFS offers more convenience and generality
for developing and maintaining applications.
Since the DFS and key-value
store have low-level APIs, it is difficult to use them for developing
applications while it is much easier and less error prone to use Odysseus/DFS
directly utilizing high-level DBMS functionality.

Second, Odysseus/DFS provides more scalability and fault-tolerance for the storage
compared to the conventional RDBMS. Expanding the storage
using a single-node RDBMS with a disk array or SAN or a distributed/parallel DBMS
has limitations in scalability, availability, and economical feasibility.
On the other hand, Odysseus/DFS overcomes these limitations by using the DFS directly
as the storage.

Third, we have shown through experiments that the performance of
Odysseus/DFS is superior to Hbase even though Odysseus/DFS provides high-level
DBMS functionality. Our experiments also show that performance degradation
of Odysseus/DFS compared to an RDBMS with local storage is not significant even though Odysseus/DFS
provides much better scalability for the storage.

For effective management of large-scale big data,
supporting both scalability of the storage and DBMS-level
functionality is important. Odysseus/DFS is the first work to
support a general purpose RDBMS for NoSQL systems. We have
shown that Odysseus/DFS is an efficient
system for processing transactions in big data.

\begin{comment}
\vspace*{-0.6cm} % orphan

\section*{Acknowledgments}

\vspace*{-0.4cm} % orphan

This work was supported by the National Research Foundation of
Korea (NRF) grant funded by Korean Government (MSIP) (No. 2012R1A2A1A05026326).

\vspace*{-0.6cm} % orphan
\end{comment}

%\input{app}


\begin{thebibliography}{10}

\bibitem{Aba09}
Abadi, D., ``Data Management in the Cloud: Limitations and Opportunities,"
{\em IEEE Data Engineering Bulletin}, Vol. 32, No. 1, pp. 3--12, Mar. 2009.

\bibitem{ABA+09}
Abouzeid, A., Bajda-Pawlikowski, K., Abadi, D., Rasin, A., and Silberschatz, A.,
``HadoopDB: An Architectural Hybrid of MapReduce and DBMS Technologies for Analytical Workloads," In {\em Proc. 35th Int'l Conf. on Very Large Data Bases (VLDB)}, pp. 922--933, Lyon, France, Aug. 2009.

\bibitem{Ale12}
Alexander, A., ``Facebook User Statistics 2012 [Infographic],'' AnsonAlex.com, Feb. 2012. available at \url{http://ansonalex.com/infographics/facebook-user-statistics-2012-infographic}.

\bibitem{BPE+10}
Blanas, S., Patel, J., Ercegovac, V., Rao, J., Shekita, E., and Tian, Y., ``A Comparison of Join Algorithms for Log Processing in MapReduce," In {\em Proc. 2010 ACM Int'l Conf. on Management of Data (SIGMOD)}, pp. 975--986, Indianapolis, Indiana, June 2010.

\bibitem{Bra08}
Brantner, M., Florescu, D., Graf, D., Kossmann, D., and Kraska, T., ``Building a database on S3," In {\em Proc. 2008 ACM Int'l Conf. on Management of Data (SIGMOD)}, pp. 251--264,  Vancouver, Canada, June 2008.

\bibitem{Bud09}
Budiu, M., presentation slides, 2009. available at \url{http://budiu.info/work/dryad-talk-berkeley09.pptx}.

\bibitem{Bur06}
Burrows, M., ``The Chubby Lock Service for Loosely-Coupled Distributed Systems," In {\em Proc. 7th USENIX Int'l Symposium on Operating Systems Design and Iplementation (OSDI)}, pp. 335--350, Seattle, Washington, Nov. 2006.

\bibitem{CDG+06}
Chang, F., Dean, J., Ghemawat, S., Hsieh, W., Wallach, D., Burrows, M., Chandra, T., Fikes, A., and Gruber, R., ``BigTable: A Distributed Storage System for Structured Data," In {\em Proc. Proc. 7th USENIX Int'l Symposium on Operating Systems Design and Iplementation (OSDI)}, pp. 205--218, Seattle, Washington, Nov. 2006.

\bibitem{CRS+08}
Cooper, B., Ramakrishnan, R., Srivastava, U., Silberstein, A., Bohannon, P., Jacobsen, H., Puz, N., Weaver, D., and Yerneni, R., ``PNUTS: Yahoo!'s Hosted Data Serving Platform," In {\em Proc. 34th Int'l Conf. on Very Large Data Bases (VLDB)}, pp. 1277--1288, Auckland, New Zealand, Aug. 2008.

\bibitem{DG04}
Dean, J. and Ghemawat, S., ``MapReduce: Simplified Data Processing on Large Clusters,"  In {\em Proc. 4th Symposium on Operating Systems Design and Implementation (OSDI)}, pp. 137--150, San Francisco, California, Dec. 2004.

\bibitem{Dig12}
The Digital Universe, EMC. available at \url{http://www.emc.com/leadership/programs/digital-universe.htm}.

\bibitem{Dit10}
Dittrich, J., Quiane-Ruiz, J., Jindal, A., Kargin, Y., Setty, V., and Schad, J.,
``Hadoop++: Making a Yellow Elephant Run Like a Cheetah (Without It Even Noticing),"
In {\em Proc. 36th Int'l Conf. on Very Large Data Bases (VLDB)}, pp. 515--529, Singapore, Sept. 2010.

\bibitem{Dit12}
Dittrich, J., Quiane-Ruiz, J., Richter, S., Schuh, S., Jindal, A., and Schad, J., ``Only
Aggressive Elephants Are Fast Elephants," In {\em Proc. 38th Int'l Conf. on Very Large Data Bases (VLDB)}, pp. 1591--1692, Istanbul, Turkey, Aug. 2012.

\bibitem{Elm10}
Elmasri, R. and Navathe, S., {\em Fundamentals of Database Systems, 6th Ed.}, Addison-Wesley, Pearson, 2010.

\bibitem{GR11}
Gantz, J., and Reinsel, D., ``Extracting Value from Chaos," {\em IDC iView}, June 2011.

\bibitem{GGL03}
Ghemawat, S., Gobioff, H., and Leung, S., ``The Google File System," In {\em Proc. 19th ACM Symposium on Operating Systems Principles(SOSP)}, pp. 29--43, Bolton Landing, New York, Oct. 2003.

\bibitem{Gra81}
Gray, J. et al., ``The Recovery Manager of the System R Database Manager," In {\em ACM Computing Surveys}, Vol. 13, pp. 223--242, 1981.

\bibitem{Har11}
Harrison, G., ``10 Things You Should Know About NoSQL Databases," TechRepublic, CBS Interactive, Feb. 2011.
available at \url{http://www.techrepublic.com/blog/10-things/10-things-you-should-know-about-nosql-databases}.

\bibitem{Hbase}
Hbase, \url{http://hbase.apache.org}.

\bibitem{HDFS}
HDFS, \url{http://hadoop.apache.org}.

\bibitem{MGL+10}
Melnik, S., Gubarev, A., Long, J., Romer, G., Shivakumar, S., Tolton, M., and Vassilakis, T., ``Dremel: Interactive Analysis of Web-Scale Datasets," In {\em Proc. 36th Int'l Conf. on Very Large Data Bases (VLDB)}, pp. 330--339, Singapore, Sept. 2010.

\bibitem{Moh92}
Mohan, C., Haderle, D., Lindsay, B., Pirahesh, H., and Schwarz, P., ``ARIES: A Transaction Recovery Method Supporting Fine-Granularity Locking and Partial Rollbacks Using Write-Ahead Logging,'' \emph{ACM Transactions on Database Systems (TODS)}, Vol. 17, No. 1, pp. 94--162, Mar. 1992.

\bibitem{ORS+08}
Olston, C., Reed, B., Srivastava, U., Kumar, R., and Tomkins, A., ``Pig Latin: A Not-So-Foreign Language for Data Processing," In {\em Proc. 2008 ACM Int'l Conf. on Management of Data (SIGMOD)}, pp. 1099--1110, Vancouver, Canada, June 2008.

\bibitem{OG10}
Ordonez C. and Garcia-Garcia, J., ``Database Systems Research on Data Mining," In {\em Proc. 2010 ACM Int'l Conf. on Management of Data (SIGMOD)}, pp. 1253--1254, Indianapolis, Indiana, June 2010.

\bibitem{Pav09}
Pavlo, A., Paulson, E., Rasin, A., Abadi, D., DeWitt, D., Madden, S., and Stonebraker, M., ``A Comparison of Approaches to Large-Scale Data Analysis," In {\em Proc. 2009 ACM Int'l Conf. on Management of Data (SIGMOD)}, pp. 165--178, Providence, Rhode Island, June 2009.

\bibitem{Shu12}
Shute, J. et al. ``F1: A Distributed SQL Database That Scales,"
In {\em Proc. 39th Int'l Conf. on Very Large Data Bases (VLDB)}, Riva del Garda, Italy, Aug. 2013.

\bibitem{Sto07}
Stonebraker, M., Madden, S., Abadi, D., Harizopoulos, S., Hachem, N., and Helland, P., ``The
End of an Architectural Era (It's Time for a Complete Rewrite)," In {\em Proc. 33rd Int'l Conf. on Very Large Data Bases (VLDB)}, pp. 1150--1160, Vienna, Austria, Sept. 2007.

\bibitem{Sto10}
Stonebraker, M., Abadi, D., DeWitt, D., Madden, S., Paulson, E., Pavlo, A., and Rasin, A., ``MapReduce and Parallel DBMSs: Friends or Foes?," {\em Communications of the ACM}, Vol. 53, No. 1, pp. 64--71, Jan. 2010.

\bibitem{TSJ+09}
Thusoo, A., Sarma, J., Jain, N., Shao, Z., Chakka, P., Anthony, S., Liu, H., Wyckoff, P., and Murthy, R., ``Hive - A Warehousing Solution Over a Map-Reduce Framework," In {\em Proc. 35th Int'l Conf. on Very Large Data Bases (VLDB)}, pp. 1626--1629, Lyon, France, Aug. 2009.

\bibitem{Val93}
Valduriez, P., ``Parallel Database Systems: Open Problems and New Issues," {\em Distributed and Parallel Databases}, Vol. 1, No. 2, pp. 137--165, 1993.

\bibitem{Waa09}
Waas, F., ``Beyond Conventional Data Warehousing --- Massively Parallel Data Processing with Greenplum Database (Invited Talk)," In {\em Book Business Intelligence for the Real-Time Enterprise}, Springer, Vol. 27, pp. 89--96, 2009.

% Whang et al. Papers

\bibitem{Wha05}
Whang, K., Lee, M., Lee, J., Kim, M., and Han, W., ``Odysseus: a
High-Performance ORDBMS Tightly-Coupled with IR Features,'' In {\em Proc. 21st
Int'l Conf. on Data Engineering}, pp. 1104--1105, Tokyo, Japan, Apr. 2005. This
paper received the Best Demonstration Award.

\bibitem{Wha12}
Whang, K., Lee, Y., Kim, Y., and Lim, H., Method for Recovering Data in a Storage System,
U.S. Patent No. 8,108,356, Jan. 31, 2012,
Application No. 12/208,014, Sept. 10, 2008.

\bibitem{Wha13}
Whang, K., Yun, T., Yeo, Y., Song, I., Kwon, H., and Kim, I.,
``ODYS: An Approach to Building a Massively-Parallel Search Engine Using a DB-IR Tightly-Integrated Parallel DBMS for Higher-Level Functionality,"
In {\em Proc. 2013 ACM Int'l Conf.
on Management of Data (SIGMOD)}, pp. 313--324, New York, New York, June 2013.

\bibitem{Wha14}
Whang, K.,  Lee, J., Lee, M., Han, W., Kim, M., and Kim, J.,
``DB-IR integration using tight-coupling in the Odysseus DBMS,''
{\em World Wide Web}, Dec. 2013, available on-line, DOI: 10.1007/s11280-013-0264-y.

\begin{comment}

\bibitem{Wha91}
Whang, K.\ and Krishnamurthy, R., ``The Multilevel Grid File - A
Dynamic Hierarchical Multidimensional File Structure,'' In {\em
Proc. Int'l Conf. on Database Systems for Advanced Applications
(DASFAA)}, Tokyo, Japan, pp. 449--459, Apr. 1991.

\bibitem{Wha02}
Whang, K., Park, B., Han, W., and Lee, Y., An Inverted Index Storage Structure
Using Subindexes and Large Objects for Tight Coupling of Information Retrieval
with Database Management Systems, U.S. Patent No. 6,349,308, Feb. 19, 2002,
Application No. 09/250,487, Feb. 15, 1999.

\bibitem{Wha03}
Whang, K., ``Tight-Coupling: A Way of Building High-Performance Application
Specific Engines,'' Presented at the panel session of {\em Int'l Conf. on
Database Systems for Advanced Applications (DASFAA)}, Tokyo, Japan, Mar. 2003,
available on-line from
http://db-www.aist-nara.ac.jp/dasfaa2003/file/Prof\_Kyu-Young\_Whang\_5.pdf.

\bibitem{Wha07a}
Whang, K., Lee, J., Kim, M., Lee, M., and Lee, K., ``Odysseus: a
High-Performance ORDBMS Tightly-Coupled with Spatial Database
Features,'' In {\em Proc. 23rd Int'l Conf. on Data Engineering},
Istanbul, Turkey, pp. 1493--1494, Apr. 2007.

\bibitem{Wha07b}
Whang, K., ``A New DBMS Architecture for DB-IR Integration,'' a keynote presentation at {\em the
APWeb/WAIM 2007 Conference}, Huang Shan, China, pp.\ 4--5, June 2007.

\bibitem{Wha09}
Whang, K., ``DB-IR Integration and Its Application to a Massively-Parallel Search Engine,'' a
keynote presentation at {\em the 18th ACM Conf. on Information and Knowledge Management (CIKM)},
Hong Kong, China, pp.\ 1--2, Nov. 2009.

\bibitem{Wha11}
Whang, K., Lee, M., Ko, H., and Kim, Y., Buffer Consistency Management Method for Multi-Server Database Management System Using the Shared-Disk Model, Korean Patent No. 10-1057468, Aug. 10, 2011, Application No. 10-2009-0072483, Aug. 6, 2009.

\end{comment}

\bibitem{Zookeeper}
Zookeeper, \url{http://zookeeper.apache.org}.

\end{thebibliography}
\end{document}